\documentclass[reprint,showpacs,amsmath,amssymb,aps,prb]{revtex4-1}
\usepackage{graphicx}
\usepackage{subfigure}
\usepackage{dcolumn}
\usepackage{bm}
\usepackage{verbatim} 
\usepackage{epstopdf}

\usepackage{amssymb}

\newcommand{\ud}{\mathrm{d}}

\begin{document}

\preprint{APS/123-QED}

\title{Order $O(1)$ algorithm for first-principles transient current through open quantum systems}

\author{King Tai Cheung}
\author{Bin Fu}
\author{Zhizhou Yu}
\author{Jian Wang}
\email{jianwang@hku.hk}
\affiliation{Department of Physics and the Center of Theoretical and Computational Physics, The University of Hong Kong, Pokfulam Road, Hong Kong, China}

\date{\today}

\begin{abstract}
In the study the response time of ultrafast transistor and peak transient current to prevent melt down of nano-chips, the first principles transient current calculation plays an essential role in nanoelectronics. The first principles calculation of transient current through nano-devices for a period of time T is known to be extremely time consuming with the best scaling $T N^3$ where N is the dimension of the device. In this work, we provide an order O(1) algorithm that reduces the computational complexity to $T^0 N^3$ for large systems. Benchmark calculation has been done on graphene nanoribbons with $N = 10^4$ confirming the O(1) scaling. This breakthrough allows us to tackle many large scale transient problems including magnetic tunneling junctions and ferroelectric tunneling junctions that cannot be touched before.
\end{abstract}

\maketitle

At the heart of growing demands for nanotechnology is the need of ultrafast transistors whose response time is one of the key performance indicators. The response of a general quantum open system can be probed by sending a step-like pulse across the system and monitored by its transient current over times, making transient dynamics a very important problem. Many experimental data show that most of the molecular device characteristics are closely related to material and chemical details of the device structure. Therefore, first principles analysis, that makes quantitative and predictive analysis of device characteristics especially its dynamic properties without relying on any phenomenological parameter, becomes a central problem of nanoelectronics.

The theoretical study of transient current dates back to twenty years ago when the exact solution in the wideband limit (WBL) was obtained by Wingreen et al.\cite{Wingreen1993a,Jauho1994}. Since then the transient current has been studied extensively using various methods\cite{Wang2013d}, including the scattering wavefunction\cite{Kurth2005,steph}, non-equilibrium Green's function (NEGF)\cite{zhu,Maciejko2006a,steph2} approach, and density matrix method\cite{chen}. The major obstacle of theoretical investigation on the first principles transient current is its computational complexity. Many attempts were made trying to speed up the calculation\cite{Kurth2005,Zhang2012,steph2,Gaury2014}. Despite of these efforts, the best algorithm to calculate the transient current from first principles going beyond WBL limit scales like $N_E T N^3$ using complex absorbing potential (CAP)\cite{Zhang2013a} where $N_E$ is a large coefficient while $T$ and $N$ are number of time steps and size of the system respectively. We note that if WBL is used, the scaling is reduced\cite{mac}. However, to capture the feature of band structure of lead and the interaction between lead and scattering region the WBL is not a good approximation in the first principles calculation.

As a result, most of the first principles investigations on transient dynamics were limited to small and simple one-dimensional systems. There are a number of problems such as magnetic tunneling junctions (MTJ)\cite{ning}, ferroelectric tunneling junctions\cite{burton}, where the system is two dimensional or even three dimensional in nature. For these systems, large number of k points $N_k$ has to be sampled in the first Brillouin to capture accurately the band structure of the system. For MTJ structure like Fe-MgO-Fe, at least $N_k = 10^4$ k points must be used to give a converged transmission coefficient\cite{hong1}. This makes the time consuming transient calculation $N_k$ times longer which is an almost impossible task even with high performance supercomputer. Clearly it is urgent to develop better algorithms to reduce the computational complexity.

In this paper, we develop a novel algorithm based on NEGF-CAP formalism to calculate transient current as a function of time step $T$. The computational time of this algorithm is independent of $T$ and therefore order $O(1)$. Four important ingredients are essential to achieve this : (1). the availability of exact solution of transient current based on non-equilibrium Green's function (NEGF) that goes beyond wideband limit. (2). the use of complex absorbing potential (CAP) so that the transient current can be expressed in terms of poles of Green's function. (3). within NEGF-CAP formalism the transient current can be calculated separately in space and time domain making $O(1)$ algorithm possible. At this point the computational complexity reduces to $50 N^3 + T N^2$. (4). the exploitation of Vandermonde matrix enables us to use fast multipole method\cite{fmm1} and fast Fourier transform to further reduce the scaling to $50 N^3 + 2 N^2 \log_2 N$ for $T<N^2$ and large $N$, therefore completely independent of $T$. To verify the computational complexity, we carry out benchmark calculations on graphene nanoribbons using the tight-binding model. A speed up factor of $1000T$ is gained for a system size of $N=2400$. A calculation is also done for the same system with $N=10200$ confirming the $O(1)$ scaling. This fast algorithm makes the computational complexity of first principles transient current calculation comparable to that of static calculation. The huge speed gain allows one to perform first principles transient calculation on a modest workstation.



For a general open quantum system with multiple leads under a step-like bias pulse, the Hamiltonian is given by
\begin{eqnarray}
H&=&\sum_{k\alpha} \epsilon_{k\alpha}\hat{c}_{k\alpha}^\dag \hat{c}_{k\alpha}
+ \sum_{n} (\epsilon_{n}+U_n(t))\hat{d}_{n}^\dag \hat{d}_{n} \nonumber \\
&+&\sum_{k\alpha n}h_{k\alpha n}\hat{c}_{k\alpha}^\dag \hat{d}_{n}+c.c. \nonumber
\end{eqnarray}
where $c^\dag$ ($c$) denotes the electron creation (annihilation) operator in the lead region. The first term in this equation corresponds to the Hamiltonian of leads with $\epsilon_{k_\alpha}$ the energy of lead $\alpha$ which contains external bias voltage $v_\alpha(t)=V_\alpha \theta(t)$. The second and third terms represent the Hamiltonian in the central scattering region and its coupling to leads, respectively. Here we have included the time-dependent Coulomb interaction $U_n(t)$ in the scattering region. This Hamiltonian can be obtained using first-principle method or assumed to be a tight-binding form. The time-dependent terminal current $I_\alpha(t)$ of lead $\alpha$ is defined as\cite{Zhang2013a}
\begin{eqnarray}\label{I}
I_\alpha(t)= \mathrm{2ReTr}[\overline{\Gamma}_\alpha H G^<(t,t)\overline{\Gamma}_\alpha-i\overline{\Gamma}_\alpha\partial_t G^<(t,t)\overline{\Gamma}_\alpha]
\end{eqnarray}
where ${\overline{\Gamma}_\alpha}$ is an auxiliary projection matrix
which is used for measuring the transient current passing through the lead $\alpha$.
Here $G^{<}$ and $H$ are the lesser Green's function and the Hamiltonian of the central scattering region, respectively.
An exact solution for $G_{CC}^<$ has been obtained by Maciejko et al\cite{Maciejko2006a} which goes beyond the WBL and has been applied to first principles calculation of transient current for atomic junctions\cite{foot1}. In terms of spectral function $A_\alpha(\epsilon, t)$, the lesser Green's function $G^<$ is given by\cite{Maciejko2006a}
\begin{align}
G^{<}(t,t)=& i\sum_\alpha \int \frac{\ud\epsilon}{2\pi}f(\epsilon) A_\alpha(\epsilon,t)\Gamma_\alpha(\epsilon)A^\dagger_\alpha(\epsilon,t) \label{eq4}
\end{align}
If we consider the upward step-like bias pulse the $A_\alpha(\epsilon,t)$ is found to be\cite{Maciejko2006a}
\begin{eqnarray}\label{Aslf}
A_\alpha(\epsilon,t)&=&\overline{G}^r(\epsilon+\Delta_\alpha)-
\int\frac{\ud\omega}{2\pi i}\frac{e^{-i(\omega-\epsilon)t}\overline{G}^r(\omega+\Delta_\alpha)}{(\omega-\epsilon+\Delta_\alpha-i0^+)}\times \notag\\
&~&\left[\frac{\Delta_\alpha}{(\omega-\epsilon-i0^+)}+\Delta \tilde{G}^r(\epsilon) \right]\notag\\
&~&\equiv A_{1\alpha}(\epsilon+\Delta_\alpha)+\int d\omega e^{-i(\omega-\epsilon) t} A_{2\alpha}(\omega,\epsilon)
\end{eqnarray}
where $\Delta_\alpha$ is the amplitude of external bias $-qV_\alpha$, $U(t)=U_{eq}+\Delta \theta(t)$ describes the potential landscape in the scattering region and $\Delta= U_{neq} - U_{eq}$ is a matrix where the subscript 'neq' and 'eq' refer to non-equilibrium and equilibrium potentials, respectively.

Despite the simplification from the conventional double time $G^<(t,t')$ to single time $G^<(t,t)$ used in Eq.(\ref{I}), the computational cost to obtain $G^<$ remains very demanding due to the following reasons. (1) Consider $A_\alpha(\epsilon,t)$ with a matrix size of $N$, matrix multiplications $\overline{G}^r(\omega+\Delta_\alpha)$ and $\tilde{G}^r(\epsilon)$ in the integrand of Eq.(\ref{Aslf}) requires computational complexity of $O(N^3)$ for each time step. As a result, the total computational cost over a period of time is at least $O(T N^3)$ where $T$ is the number of time steps. (2) Double integrations in energy space are required for $G^<$. The presence of numerous quasi-resonant states whose energies are close to real energy axis makes the energy integration in $A_\alpha$ extremely difficult to converge.
This problem can be overcome using the complex absorbing potential (CAP) method\cite{Driscoll2008}.
The essence of CAP method is to replace each semi-infinite lead by a finite region of CAP while keeping transmission coefficient of the system unchanged. In addition, it has been demonstrated in Ref.\onlinecite{Zhang2013a} that the first principles result of transient current for molecular junctions obtained from the exact numerical method (non-WBL) and the CAP method are exactly the same. Using the CAP method, the poles of the Green's function can be obtained easily and the spectral function can be calculated analytically using the residue theorem. Expanding Fermi function using Pade approximant (PSD)\cite{Hu2011} further allows us to calculate the transient current separately in space and time domain making $O(T^0 N^3)$ algorithm possible.

Now we illustrate how to achieve order $O(1)$ algorithm for the transient current calculation, i.e., $I_\alpha(t_j)$ for $j=1,2,...,T$. Substituting Eq.(\ref{Aslf}) to Eq.(\ref{eq4}), $G^{<}(t,t)$ can be written as
\begin{eqnarray} \label{Gless}
G^{<}(t,t)
&~&= (i/\pi)[B_1+\int d\omega d\omega' e^{-i(\omega - \omega') t} B_2(\omega,\omega') \notag\\
+&~& \sum_\alpha\int d\epsilon d\omega e^{i(\omega-\epsilon)t} f(\epsilon) A_{1\alpha} W_\alpha A^\dagger_{2\alpha}+c.c.]
\end{eqnarray}
where $B_{1}=\int d\epsilon f(\epsilon) \sum_\alpha A_{1\alpha} W_\alpha A^\dagger_{1\alpha}$,
$B_{2}(\omega,\omega')=\int d\epsilon f(\epsilon) \sum_\alpha A_{2\alpha}(\omega,\epsilon) W_\alpha A^\dagger_{2\alpha}(\omega',\epsilon)$, and
$W_\alpha$ is the CAP matrix. In terms of poles of Green's function and Fermi distribution function, we have\cite{foot3}
\begin{eqnarray} \label{Gless1}
&&G^{<}(t,t)= (i/\pi)[B_1+\sum_{n m}e^{-i(\epsilon_n - \epsilon_m^*) t} {\bar B}_2(\epsilon_n,\epsilon_m^*) \notag\\
&&+ \sum_\alpha \sum_{n m}e^{-i(\epsilon_n-{\epsilon}^*_m+\Delta_\alpha)t} f({\epsilon}^*_m) {\bar B}_{3\alpha}(\epsilon_n,\epsilon_m^*)+c.c. \notag\\
&&+ \sum_\alpha \sum_{n l}e^{i({\tilde\epsilon}_l-{\epsilon}^*_n)t} {\bar f}({\tilde\epsilon}_l){\bar B}_{4\alpha}({\tilde \epsilon_l},\epsilon_n^* )+c.c.]
\end{eqnarray}
where $\epsilon_n$ ($n=1,2,...N$) is the complex energy spectrum of $H_{neq}-iW$ in the lower half plane while ${\tilde\epsilon}_l$ being the poles of $f(E)$ using PSD with $l=1,...N_f$ \cite{foot3}.

Within CAP framework, $G^<$ in Eq.(\ref{I}) is the lesser Green's function of the central scattering region excluding the CAP regions. Substituting the second term of Eq.(\ref{Gless}) into the first term in Eq.(\ref{I}), we find its contribution to current (denoted as $I_1$)
\begin{eqnarray}
I_1(t) &=& 2{\rm Re} \sum_{n m}e^{-i(\epsilon_n - \epsilon_m^*) t} {\rm Tr} [\overline{\Gamma}_\alpha H_{CC} \bar{B}_2(\epsilon_n,\epsilon_m^*)\overline{\Gamma}_\alpha] \nonumber \\
&\equiv &2{\rm Re} \sum_{n m}e^{-i(\epsilon_n - \epsilon_m^*) t} M_{nm}
\end{eqnarray}
where the matrix $M$ does not depend on time. We see that the space and time domains have been separated.

Denoting a Vandermonde matrix $V_{jk} = \exp(i \epsilon_k t_j)$ with $k=1,2,...,N$, $t_j=j dt$ , $j=1,2,...,T$ where $dt$ is the time interval, we have $I_1(t_j) = [V^t (M+M^\dagger) V^*]_{jj}$. Using this approach, we finally obtain
\begin{eqnarray}
I_\alpha(t_j) &=& I_{0\alpha}+[V^t M_1 V^*]_{jj} + ([V^t M_2 {\tilde V}^*]_{jj}+c.c.) \label{eq11}
\end{eqnarray}
where ${\tilde V}_{jk} = \exp(i{\tilde \epsilon}_k t_j)$ is a $T \times N_f$ matrix, $M_1$ is a $N \times N$ matrix while $M_2$ is a $N \times N_f$ matrix.
Since $\epsilon_k$ is the complex energy in the lower half plane, $V_{jk}$ goes to zero at large j. Hence $I_{0\alpha}$ is the long time limit of transient current which can be calculated using Landauer Buttiker formula. The time dependent part of the transient current can be separated into real space calculation (calculation of $M_1$ and $M_2$) and then a matrix multiplication involving time. We note that at room temperatures the Fermi function can be accurately approximated by 15 or 20 Pade approximants. Hence the calculation of $[V^t M_1 V^*]_{jj} + ([V^t M_2 {\tilde V}^*]_{jj}+c.c.)$ can be combined to give $T N^2$ computational complexity.

Now we examine the computational complexity. The computational complexity of real space calculation is estimated to be $50 N^3$. Therefore the total computational complexity is $50 N^3 + T N^2$. At this stage, the algorithm (denoted as algorithm I) is not $O(1)$ yet. In the supplemental material, we will show that matrix multiplication $V^t M$ can be done using fast multipole method and fast Fourier transform (denoted as algorithm II). This will reduce the computational complexity of $V^t M$ from $T N^2$ to $2 T \log_2 N$. Hence for $T<N^2$, the computational complexity is $50 N^3 + 2 N^2 \log_2 N$. For $T>N^2$, the scaling is $50 N^3 + 2 T \log_2 N$. However, for large T, the physics comes into play. Since $\epsilon_j$ is the complex energy of the resonant state, $V_{jT}=\exp(-i\epsilon_j dt T)$ decays quickly to zero before $T=N^2$. For a graphene nanoribbon with $N=10^4$ (see details below), the maximum value of $V_{jT}=\exp(-i\epsilon_j dt T)$ is $10^{-3}$ when $T=N$ and $dt = 0.1$ fs. Consequently all the matrix elements are zero for $T=10N$. Hence for large systems, there is no need to go beyond $T=N^2$. In this sense, the algorithm II is order $O(1)$ algorithm.

\begin{figure}[h]
\centering
\includegraphics[width=0.5\textwidth]{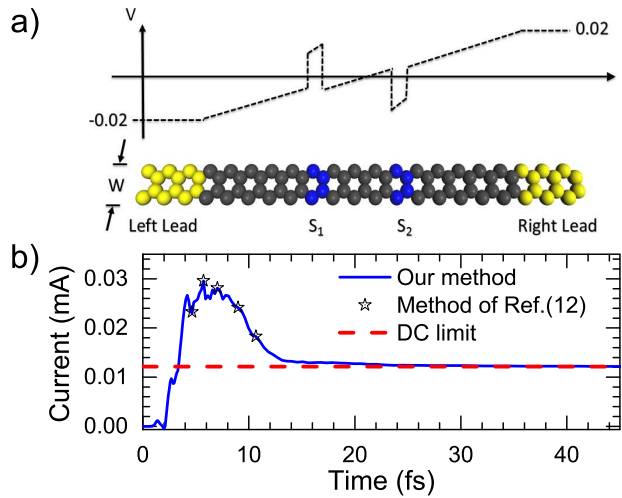}
\caption{a) Configuration of the gated graphene nanoribbon. The $D_1$ and $D_2$ gate are of values 0.03V and -0.03V respectively. b) Transient current of zigzag graphene nanoribbon for a system of 600 atoms The dashed line is the dc limit.}
\label{fig1}
\end{figure}
To demonstrate the power of this algorithm, we calculate the transient current in a graphene nanoribbon.
Graphene is a well-known intrinsic 2D material with many exotic properties\cite{CastroNeto2009,Zhang2005}. Its transient behaviour in response to a step-like pulse was studied in the literature\cite{steph,Perfetto2010,Klymenko2008}. We will test our algorithm on a gated graphene nanoribbon at room temperature using the tight-binding (TB) Hamiltonian given by:
\begin{align}
\hat{H}=-h \sum_{<i,j>} \hat{c}^\dag_i \hat{c}_j  - q\sum_{i} [V_i \theta(t)+V_{g1i}+V_{g2i}] \hat{c}^\dag_{i} \hat{c}_{i}
\end{align}
where $\hat{c}^\dag_i$ ($\hat{c}_i$) is the creation (annihilation) operator at site i and $h=2.7$eV being the nearest hopping constant. Here $V(x) = V_L + (V_R-V_L)x/L$ is the potential landscape due to the external bias with $V_R=-V_L=0.02$V and $V_{g1}$ and $V_{g2}$ are gate voltages in regions $D_1$ and $D_2$, respectively.

We first confirm that the transient current calculated using the new method is the same as that of Ref.(\onlinecite{Zhang2013a}). Using 30 layers of CAP, transmission coefficient versus energy was calculated which shows good agreement with the exact solution. This also ensures the correct steady state current. For the transient current, excellent agreement is also obtained between our algorithm and that of Ref.(\onlinecite{Zhang2013a}) (see Fig.(\ref{fig1})).
We note that with the introduction of gates, the 'on-off' time of graphene is shortened in comparison to un-gated graphene which has a long oscillating current\cite{steph}.

\begin{figure}[h]
\centering
\includegraphics[width=0.5\textwidth]{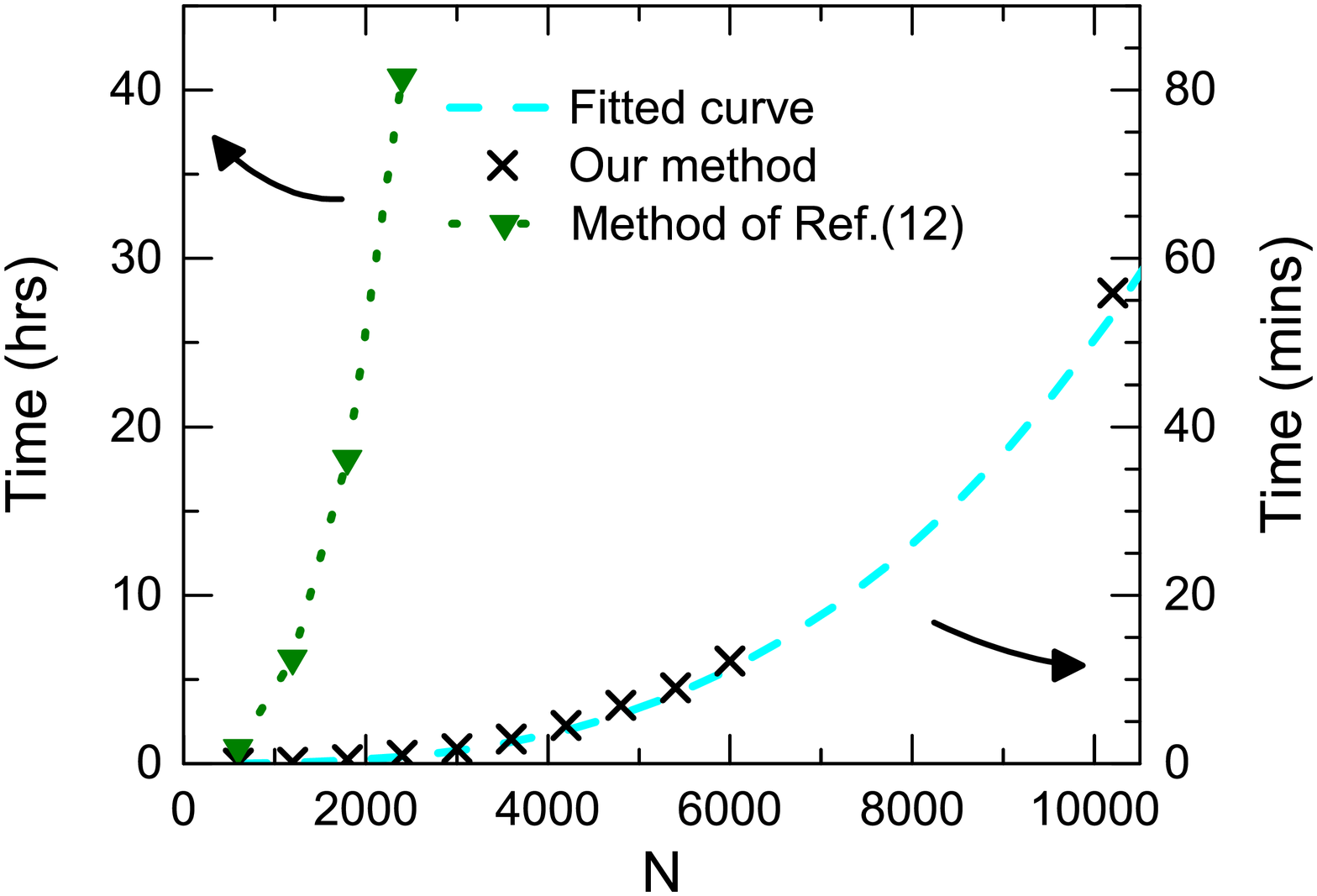}
\caption{Scaling of N against computation time at T=3. The fitted curve in the form of $50N^3+TN^2$ is in good agreement with
the calculated results (Y-axis is on the right). In order compare the performance of Ref.(\onlinecite{Zhang2013a}), 6000 energy points was used for integration(Y-axis is on the left).}
\label{fig2}
\end{figure}

\begin{figure}
\centering
\includegraphics[width=0.5\textwidth]{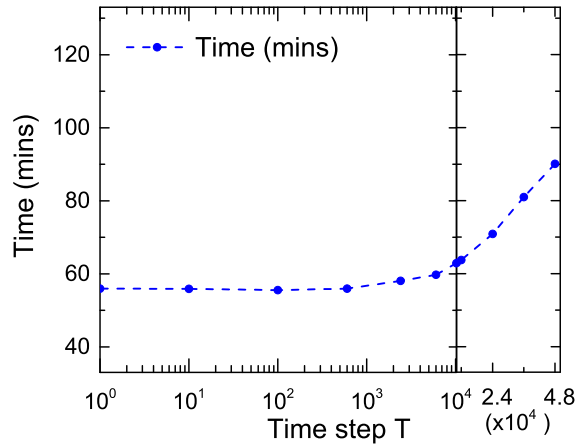}
\caption{Scaling of T against computational time for N=10200. Left hand side: exponential scale in T; Right hand side: linear scale in T, shows that at extreme large data points ranges over 10 thousands points, computational time is proportional to T.}
\label{fig3}
\end{figure}

Now we test the scaling of our algorithm by calculating the transient current for nanoribbons with different system sizes ranging from 600 to 10200 atoms\cite{foot2}. We first test the algorithm I. Computational time of transient current for 3 time steps against system sizes N is shown in Fig. (\ref{fig2}). We have fitted the data using $50N^3+TN^2$ with very good agreement showing $T N^2$ scaling for the time-dependent part. For comparison, we have also plotted the computation time using method in Ref.(\onlinecite{Zhang2013a}). We found that the number of energy points $N_E$ depends on the spectrum of resonant states of the system. For graphene nanoribbons with 600 atoms, we have used $N_E=6000$ to converge the integral over Fermi function. Fig. (\ref{fig2}) shows that a speed up factor of 1000T is achieved at $N=2400$. The scaling is shown in Fig. (\ref{fig3}), from which we see that for $T<N$ the computational time is almost independent of the number of time steps.

Now we examine the algorithm II which reduces the scaling $T N^2$ further. Notice that the scaling $T N^2$ comes from matrix multiplication involving Vandermonde matrix $V^t M_1$. Fast algorithm is available to speed up the calculation involving structured matrix such as Vandermonde matrix. As discussed in details in the supplemental material, we can use fast multipole method\cite{fmm1} and fast Fourier transform (FFT) to carry out the same matrix multiplication using only $c_3 N^2 \log_2 N$ operations provided $T<N^2$.  Here the coefficient $c_3$ is a large constant that depends only on the tolerance of the calculation and the setup of fast multipole method (FMM). The numerical calculation using our FMM code shows that for $T=N=10^4$, the FMM together with FFT already outperforms the $T N^2$ scaling by a factor of 8. Of course, there are lots of room to optimize the FMM calculation.

\bigskip

We now discuss several fast algorithms proposed recently to calculate transient current. Assuming WBL approximation, a scaling of $T N^3$ was achieved for the transient current at zero temperature by Tuovinen et al \cite{steph2} and this scaling was recently reduced to $T N^2$ by Ridley et al\cite{mac}. We note that these algorithms cannot be used for first principles transient current calculations since it assumes WBL approximation. Recently, an algorithm of numerically solving time dependent Schrodinger equation {\it explicitly} has been proposed\cite{waintal}. The scaling of obtaining the scattering wavefunction for a given energy is $T N$ which translates to $T N_E N$ for the transient current. However, we note that the {\it implicit} scheme must be used for a stable solution of time dependent Schrodinger equation which scales at least $T N_E N^3$ as shown in Ref.\onlinecite{Kurth2005}. On the other hand, our algorithm is stable and goes beyond WBL suitable for the first principles calculation.

Since our algorithm is based on the NEGF-CAP formalism, it can easily be extended to the NEGF-DFT-CAP formalism which performs the first principles calculation. In fact, the NEGF-DFT-CAP method has already been successfully implemented in the first principles transient current calculation in Ref.(\onlinecite{Zhang2013a}) which gives exactly the same result from the NEGF-DFT.
With the order $O(1)$ algorithm at hand, many applications can be envisaged. For instance, the transient spin current
(related to spin transfer torque) using the NEGF-DFT-CAP formalism has been carried out for planar structures
where k-sampling in the first Brillouin zone is needed.
Our $O(1)$ method can include k-sampling easily. It is also straightforward to obtain exact solution of transient current by
including electron-phonon interaction in the Born approximation as well as other dephasing mechanism. Finally, first principles transient photo-induced current on two dimensional layered materials can be calculated using our method.

\bigskip

\begin{acknowledgements}
This work was financially supported by the Innovation and Technology Commission of the HKSAR (ITS/217/14), the University Grant Council (Contract No. AoE/P-04/08) of the Government of HKSAR, and NSF-China under Grant No. 11374246.
\end{acknowledgements}
\newpage

\section{Appendix}

{\noindent {\bf Pade approximant}}

Brute force integration over Fermi function along real energy axis to obtain $G^<(t,t)$ may need thousands of energy points to converge which is very inefficient. To obtain an accurate result while reducing the cost, fast converging PSD is used for the Fermi function $f$ in eq.(\ref{Gless}) so that the residue theorem can be applied.\\

Using [n-1/n] PSD scheme\cite{Hu2011}  with the Pade approximant accurate up to $O((\epsilon/kT)^{4n-1})$, Fermi function $f$ can be expressed as
\begin{align}\label{fermi}
f(\epsilon)=\dfrac{1}{2}-\sum_{j=1}^n\dfrac{2\eta_j\beta \epsilon}{(\beta \epsilon)^2+{\xi_j}^2}
\end{align}
where $\xi_j$ and $\eta_j$ are two set of constants that can be calculated easily. Using the PSD scheme analytic form of $G^<$ in eq.(\ref{Gless}) can be obtained using the residue theorem.

\bigskip

{\noindent {\bf Calculation of spectral function}}

We express $\tilde{G}^{r}(\epsilon)$ and $\overline{G}^{r}(\epsilon)$, the equilibrium and non-equilibrium retarded Green's functions, respectively in terms of their eigen-functions by solving the following eigen-equations for $H_{eq}$ and $H_{neq}$,\cite{Zhang2013a}
\begin{eqnarray}
(H_{eq} -iW) \psi^0_n &=& \epsilon_n^0 \psi^0_n \nonumber \\
(H_{eq} +iW) \phi^0_n &=& \epsilon_n^0 \phi^0_n
\end{eqnarray}
where $W = \sum_\alpha W_\alpha$ and similar equations can be defined for $H_{neq}$.
Using the eigen-functions of $H_{eq}+iW$ and $H_{neq}+iW$, we have
\begin{align}
\tilde{G}^{r}(\epsilon)&=[\epsilon - H_{eq}+iW]^{-1}=\sum_n \frac{|\psi^0_n\rangle\langle\phi^0_n|}{(\epsilon-\epsilon^0_n+i0^{+})}\label{Geq},\\
\overline{G}^{r}(\epsilon)&=[\epsilon - H_{neq}+iW]^{-1}=\sum_n \frac{|\psi_n\rangle\langle\phi_n|}{(\epsilon-\epsilon_n+i0^{+})}\label{Gneq}.
\end{align}
Performing integral over $\omega$ using the residue theorem, the analytic solution of $A_\alpha$ is obtained
\begin{align}\label{A}
A_\alpha(\epsilon,t)
=&\sum_n \frac{|\psi_n\rangle\langle\phi_n|}{\epsilon+\Delta_\alpha-\epsilon_n+i0^+}+
\sum_n\frac{e^{i(\epsilon+\Delta_\alpha-\epsilon_n)t}|\psi_n\rangle\langle\phi_n|}{\epsilon-\epsilon_n+i0^+}\times\notag\\
&\left[\frac{\Delta_\alpha}{\epsilon+\Delta_\alpha-\epsilon_n+i0^+}-\Delta\sum_l\frac{|\psi^0_l\rangle\langle|\phi^0_l|}{\epsilon-\epsilon^0_l+i0^+}\right],
\end{align}
where $\Delta=H_{neq}-H_{eq}$.

\bigskip

{\noindent {\bf Calculation of lesser Green's function}}

In Eq.(\ref{Gless1}), ${\bar B}_2$ is defined as

$${\bar B}_2 = -4\pi^2 \left[B_2(\omega,\omega')(\omega-\epsilon_n)(\omega'-\epsilon_m^*)\right]|_{\omega=\epsilon_n,\omega'=\epsilon^*_m}$$

and

$${\bar B}_{3\alpha} = -4\pi^2 \left[ A_{1\alpha}({\epsilon}) W_\alpha A^\dagger_{2\alpha}(\epsilon,\omega)(\epsilon-\epsilon_n)(\omega-\epsilon_m^*)\right]|_{\epsilon=\epsilon_n,\omega=\epsilon^*_m}$$
and
$${\bar B}_{4\alpha}=2\pi i  A_{1\alpha}({\tilde\epsilon}_l) W_\alpha \left[ A^\dagger_{2\alpha}({\tilde \epsilon}_l,\omega)(\omega -\epsilon_n^*) \right] |_{\omega=\epsilon^*_n}$$
and $${\bar f} = 2\pi i (f(\epsilon)(\epsilon-{\tilde\epsilon}_l))|_{\epsilon={\tilde\epsilon}_l} $$
\bigskip

{\noindent {\bf Fast multipole method}}

The fast multipole method\cite{fmm1} has been widely used and has been ranked top 10 best algorithms in 20th Century\cite{fmm}. It is extremely efficient for large N. We want to calculate the following quantity
\begin{equation}
\label{a1}
I(t)=\sum_{n,m} \exp(-i\epsilon_nt) M_{nm} \exp(i\epsilon_m^* t)
\end{equation}
where the matrix $M$ can be expressed in terms of vectors as $M=(c_0,c_1,...,c_{N-1})$ and $V_{n j} = \exp(-i\epsilon_n t_j)$ is a Vandermonde matrix with $t_j=jdt$ and $j=1,2,...T$. Eq.(\ref{a1}) is of the form
$V^t M V^*$ where $t$ stands for transpose. In the following, we outline how to calculate $V^t c$ where $c$ is a vector of $N$ components.

Setting $a_j = \exp(-i\epsilon_j)$ and denoting $T$ the number of time steps. Then $b=V^t c$ is equivalent to $b_n = \sum_{j=0}^{N-1} c_j a_j^n$. A direct computation shows that the entries of $b=V^t c$ are the first $T$ coefficients of the Taylor expansion of
\begin{equation}
S(x)=\sum_{j=0}^{N-1} \frac{c_j}{1-a_j x}= \sum_n^\infty \sum_{j=0}^{N-1} c_j \left( a_j x \right)^n = \sum_n b_n x^n \label{a4}
\end{equation}
where $b_n = \sum_{j=0}^{N-1} c_j (a_j)^n$. Denoting ${\bar S}(x)=\sum_{n=0}^{T-1} b_n x^n$ and setting $x=(\omega_{T})^l$ with $\omega_{T}=\exp(i 2\pi/T)$ we can calculate $\bar{S} (\omega_{T}^l)$ which is the Fourier transform of $b_n$,
\begin{eqnarray}
\bar{S} (\omega_{T}^l) &=& \sum_{j=0}^{N-1} \sum_{n=0}^{T-1} c_j (a_j)^n \omega_{T}^{nl} =\sum_{j=0}^{N-1} c_j \frac{1-\left( a_j \omega_{T}^l \right)^{T }}{1-a_j \omega_{T}^l} \nonumber \\
                     &=& \omega_{T}^{-l}\sum_{j=0}^{N-1} \frac{c_j (1-a_j^T)}{{\left( 1/\omega_{T}\right) }^l-a_j }  \nonumber
\end{eqnarray}
where we have used $\omega_{T}^T=1$. Note that the fast multipole method (FMM) aims to calculate $v_l = \sum_j c_j/(x_l-a_j)$ with $O(N)$ operations instead of $N^2$ operations. Hence $\bar{S} (\omega_{T}^l)$ can be obtained using FMM, from which we calculate $b_n$ using FFT.

Now we estimate the computational complexity for $T \leq N$. For FMM we need $\kappa_1 {\rm max} (T,N)$ operations where $\kappa_1$ is about $40 \log_2(1/\tau)$ with $\tau$ the tolerance. For FFT the computational complexity is at most $\kappa_2 N \log_2 N$ where $\kappa_2$ is a coefficient for FFT calculation. To compute $V^t M$ where $M$ has $N$ vectors, we have to calculate $V^t c$ $N$ times. Hence the total computational complexity is $\kappa_1 N^2 + \kappa_2 N^2 \log_2 N$. For $T=N=10^4$, numerical calculation using FMM and FFT shows that $\kappa_1 N^2$ dominates due to large $\kappa_1$ and the speed up factor is about 8 over $T N^2$ scaling discussed in the main text.

For very large $T$ up to $T=N^2$ (if $N=10^4$ we have $T=10^8$), we will show that the computational complexity is $\kappa_1 N^2 + 2\kappa_2 N^2 \log_2 N $. In fact, it is easy to see that $I(t_j)$ defined in Eq.(\ref{a1}) is the first $T$ coefficients of the Taylor expansion of
\begin{eqnarray}
S(x)&=&\sum_{n,m=0}^{N-1} \frac{M_{nm}}{1-a_n a_m^* x} \label{eq17}\\
&=& \sum_j^\infty \sum_{n,m=0}^{N-1} M_{nm} \left( a_n a_m^*\right) ^j x^j = \sum_j I(t_j) x^j
\end{eqnarray}
where $a_n = \exp(-i\epsilon_n)$. Now we define two new vectors $u$ and $d$ which have $N^2$ components with $u^t =(c_0^t, c_1^t, ..., c_{N-1}^t)$  (recall our definition $M=(c_0,c_1,...,c_{N-1})$) and $d^t = (a_0^* a^t, a_1^* a^t, ..., a_{N-1}^* a^t)$, where once again t stands for transpose. With the new vectors defined, $S(x)$ in Eq.(\ref{eq17}) is expressed as
\begin{eqnarray}
S(x)=\sum_{j=0}^{N^2-1} \frac{u_j}{1-d_j x} \label{a5}
\end{eqnarray}
which is exactly the same form as Eq.(\ref{a4}). The only difference is that $c$ and $a$ in Eq.(\ref{a4}) have $N$ components and $\tilde S$ has to be calculated $N$ times while $u$ and $d$ in Eq.(\ref{a5}) have $N^2$ components and we calculate $\tilde S$ defined according to Eq.(\ref{a5}) just once. Therefore the computational complexity is $\kappa_1 N^2 + \kappa_2 N^2 \log_2 N^2 $. If $T=nN$ with $n=1,2,...N$, it is not difficult to show that the computational complexity is $\kappa_1 T N/n + \kappa_2 T (N/n) \log_2 (nN) =\kappa_1 N^2 + \kappa_2 N^2 \log_2 (nN) $.

To summarize, the computational complexity of Eq.(\ref{a1}) is $\kappa_1 N^2 + 2\kappa_2 N^2 \log_2 N $ for $T<N^2$.
It is easy to show that for $T>N^2$ the scaling is $\kappa_1 N^2 + 2 \kappa_2 T \log_2 N$. However, for large T, the physics comes into play. Since $a_j=\exp(-i\epsilon_j)$ with $\epsilon_j$ the energy of resonant state, $a_j^T$ quickly decays to zero before $T=N^2$ and hence no need to go up for $T>N^2$.


\begin{thebibliography}{34}
\bibitem{foot2}
The channel length is fixed at $L = 11.07$ nm including the CAP regions while
the width changes from 2 unit cells (0.85 nm) to 34 unit cells (14.5nm). We have used 16 terms for Pade approximants.
\end{thebibliography}

\end{document}